# MCell-R: A particle-resolution network-free spatial modeling framework

Jose-Juan Tapia, Ali Sinan Saglam, Jacob Czech, Robert Kuczewski, Thomas M. Bartol, Terrence J. Sejnowski, and James R. Faeder

**Running Head:** MCell-R

**Corresponding Author:**

James R. Faeder

Department of Computational and Systems Biology

University of Pittsburgh

Pittsburgh, PA 15260, USA

Email: faeder@pitt.edu

**Summary:**

Spatial heterogeneity can have dramatic effects on the biochemical networks that drive cell regulation and decision-making. For this reason, a number of methods have been developed to model spatial heterogeneity and incorporated into widely used modeling platforms. Unfortunately, the standard approaches for specifying and simulating chemical reaction networks become untenable when dealing with multi-state, multi-component systems that are characterized by combinatorial complexity. To address this issue, we developed MCell-R, a framework that extends the particle-based spatial Monte Carlo simulator, MCell, with the rule-based model specification and simulation capabilities provided by BioNetGen and NFsim. The BioNetGen syntax enables the specification of biomolecules as structured objects whose components can have different internal states that represent such features as covalent modification and conformation and which can bind components of other molecules to form molecular complexes. The network-free simulation algorithm used by NFsim enables efficient simulation of rule-based models even when the size of the network implied by the biochemical rules is too large to enumerate explicitly, which frequently occurs in detailed models of biochemical signaling. The result is a framework that can efficiently simulate systems characterized by combinatorial complexity at the level of spatially-resolved individual molecules over biologically relevant time and length scales.

## 1. Introduction

Computational modeling has become an important tool for studying the dynamics of complex reaction networks [1]. In traditional modeling approaches a modeler defines the species of interest together with a reaction network that specifies the kinetics of the system. If the number of individual reactant molecules in the system is high ($10^3$ or greater), it is possible to simulate the model deterministically by numerically solving ordinary differential equations (ODEs). But if the number of individual molecules is on the order of hundreds or smaller, then stochastic effects may be important and stochastic simulation methods, such as Gillespie's stochastic simulation algorithm (SSA) [2], are required. A number of software platforms, such as CellDesigner [3] and COPASI [4], facility the specification, simulation, and visualization of reaction networks and their dynamics.

Another important consideration in the development of accurate models is the potential importance of spatial effects. Both the ODE and SSA approaches assume that the involved molecules can be treated as if they are uniformly distributed in space. That such models often give accurate representation of observed dynamics and meaningful predictions is belied by the fact that cells are anything but well-mixed containers. In fact, spatial models are required to accurately describe many biochemical phenomena at the cell level, including transmission of signals in neuronal spines [5], organization of proteins that control cell division in bacteria [6], and many other systems [7], [8]. Spatial considerations can be added to a reaction network model by defining compartments and compartment boundaries that restrict the movement of molecules in a system. The spatial description can be further refined by defining compartment geometries and explicitly representing the species concentrations as a function of position. If species concentrations are taken to be continuous, the system can be modeled as a set of partial differential equations (PDE) that are affected by both the diffusion of the species and their reactions with each other. The Virtual Cell is an example of a simulation platform that provides a general purpose implementation of this approach for cell biological models [9]. Spatial simulations feature two distinct approaches to treating stochastic effects arising from discrete molecular populations. At the so-called "mesoscopic" level, space is divided into a set of voxels, each of which tracks the number of each species it contains [10], [11]. StochSS is a general purpose simulation tool that implements this approach [12]. Diffusion between voxels and reactions within voxels are tracked as discrete events, but individual particles are not tracked. At a finer level of resolution, particle-based methods instantiate every molecule in the system and model their diffusion and reaction explicitly. Smoldyn [13] and MCell [14] enable the development and simulation of such models.

A common thread connecting these approaches is that the modeler must define the full reaction network as part of the model specification. Biochemical networks that regulate cellular function are often characterized by combinatorial complexity, which can make manual specification of the reaction network tedious or even infeasible [15], [16]. For example, a receptor with 10 phosphorylation sites has $2^{10} = 1024$ states of phosphorylation and a correspondingly large number of possible reactions. Aggregation of receptors or binding of adaptor molecules to these sites can create complexes with an astronomical number of possible states. Rule-based modeling is a paradigm that was developed to deal with combinatorial complexity by building up species and reaction networks from structured molecules and rules [17]–[19]. Its graph-based approach to model specification allows the full reaction network to be specified by a much smaller number of reaction rules [20]. The modeling frameworks BioNetGen [21], Kappa [22], and Simmune [23] are examples of software frameworks that that implement this rule-based approach.

A number of spatial simulators have integrated rule-based modeling capabilities. Simmune [23] uses a subvolume-based PDE approach for reaction dynamics such that the necessary equations are generated on-the-fly within each subvolume based on local concentrations and the global set of rules. This approach, however, neglects stochastic effects. BioNetGen has a compartmental extension that considers the division of the system into well mixed-subvolumes [24], which allows for both deterministic and stochastic simulations, but provides a lower degree of spatial resolution. The Stochastic Simulation Compiler [25] combines a mesoscopic stochastic spatial approach with rule-based model definition and pre-compiles the expanded reaction-network into assembly-language for efficient simulation. Unfortunately, the software is no longer actively maintained or developed, and works on a dwindling number of platforms, not including Microsoft Windows. SpatialKappa [26] is an extension to the Kappa language and simulation tools that also implements next-subvolume diffusion. Smoldyn was extended to incorporate rule-based modeling capabilities based on either a wild-card based syntax (***see Andrews, Chapter 8***) or BioNetGen language [27]. SpringSaLaD [28] performs Brownian Dynamics simulations, which unlike other spatial simulators mentioned so far include the effects of volume exclusion, based on a multi-state multi-component specification. The drawback of this approach, however, is that it requires a much smaller timestep compared with other spatial simulators, which makes it impractical for simulations on the cellular length and time scales. Another platform that integrates a Brownian Dynamics simulator with rule-based model specification is SRsim [29], [30].

One limiting factor for most of these spatial simulators is the need for the reaction-network to be generated from the rule-specification prior to simulation. Even though the rule-based approach facilitates the specification of a model, in the face of a high degree of combinatorial complexity pre-computation may become a liability [16], [31]. For example, it was shown that generating the full reaction network for a model of the CaMKII system on a standard 2.54GHz Intel Xeon processor would take 290 years [32].

The basic premise of network-free simulators is to individually store in memory every molecular species in the system as an independent object, such that their progress is tracked throughout the course of the simulation. The algorithm then proceeds to directly map the set of reaction rules in the system (instead of the full reaction network) to these particle agents whenever a biological event is scheduled to occur. If an event is triggered, then a set of matching particles is chosen as reactants and transformed to create the products specified by the reaction rule. This approach avoids the need to pre-compute the full reaction network at the cost of keeping the complete set of molecular agents in memory. The memory cost of this network-free approach scales linearly with the number of rules and particles instead of the number of possible species and reactions. Since the number of rules is typically much lower than the number of reactions, there can be a substantial memory savings [33]. Some examples of non-spatial simulation platforms that implement a network-free approach are StochSim [34], RuleMonkey [35], NFsim [16], and KaSim [22].

This chapter presents a new rule-based, spatial modeling framework that provides accurate simulation results at the particle resolution scale and that is not limited by combinatorial complexity in its simulation efficiency. The simulator we have developed, MCell-R, integrates two existing simulators that we have mentioned above: MCell and NFsim. The MCell spatial simulation engine provides efficient simulation of particle-based reaction-diffusion dynamics in arbitrarily complex geometries, and NFsim provides a library of functions to carry out the graph operations required for efficient network-free simulation in a spatial context. As described in ***Section 2, Materials***, MCell's particle-based simulation algorithm has been extended to use NFsim to determine reaction probabilities involving arbitrarily complex multi-state and multi-component species. This integrated capability allows models to be simulated efficiently regardless of the size of the reaction network implied by the rules and without generating the full network. A user is thus free to explore the effects of features such as multi-site phosphorylation and multivalent binding without having to worry about the number of possible species and reactions.

***Section 3, Methods***, introduces a language extension that we have developed for MCell's Model Description Language (MDL) that we have called MDL rules or MDLr for short. This extension incorporates rule-based descriptions of structure molecules and reaction rules based on the BioNetGen language (BNGL) syntax. Several examples are provided along with validation tests that demonstrate the accuracy of the MCell-R simulator.

## 2. Materials

At the time of this writing, MCell-R is currently under active development. Source code and instructions for compiling MDLr can be found at the GitHub repository for MCell (http://github.com/mcellteam/mcell). We also plan to release installation packages for major platforms including MacOS, Linux, and Windows, which will be available on the MCell web site (http://mcell.org). Work is also underway to incorporate MCell-R into CellBlender, which is our graphical interface for spatial modeling that enables interactive specification and simulation of spatial models without writing MDL files [36]. The main goal of this section is to present the algorithmic and software extensions to both NFsim and MCell that were required for the development of MCell-R. The section is divided into two parts with ***Sections 2.1*** and ***2.2*** describing extensions to NFsim and MCell respectively that were required to construct the integrated simulator.

Currently, MCell-R models are specified in the MDLr language that will be presented in more detail in ***Section 3, Methods***. MDLr is an extension of MDL that allows the user to introduce multi-state multi-component elements, including molecule components and states as well as reaction rules that operate on these into an MCell model definition. As shown in ***Fig.* 1,** the MDLr preprocessor extracts the rule-based graph information from an MDLr input file and generates two separate input files: an MCell model spatial definition encoded in MDL and rule-based model definition that is used for NFsim initialization encoded in BNGL. These model definitions are then used to initialize MCell and NFsim components separately. In order to facilitate efficient simulation, NFsim functionality has been encapsulated in a software library that is invoked by MCell at runtime so that the simulator runs in a single process. To construct the integrated MCell-R simulator, the following extensions were developed to the NFsim and MCell frameworks:

**NFsim**

- Encapsulation of the functionality present in NFsim as a stand-alone API such that it can be incorporated in other simulation frameworks (***Section 2.1.1***).
- Implementation of the compartmental BioNetGen specification in NFsim such that spatial considerations can be taken into account during the selection of graph-based events (***Section 2.1.2***).
- Development of methods to determine diffusion constants for complexes composed of multiple molecules (***Section 2.1.3***).
- Implementation of a hierarchical namespace framework such that attributes and properties can be attached to complexes, compartments, molecule types, reaction rules (***Section 2.1.4***).

**MCell**

- Extensions to the MCell event scheduler such that it can handle network-free events, structured objects and their properties by communicating with NFsim (***Section 2.2.1***).
- Extensions to the MDL specification language and the MCell internal model representation to define, initialize and track multi-state, multi-component objects (***Section 2.2.2***).

The remainder of this section describes these extensions in further detail. Readers who are primarily interested in using MCell-R may safely skip to the tutorial provided in ***Section 3***.

### 2.1 Extensions to NFsim

#### *2.1.1 Development of* libNFsim

In order to leverage the simulation capabilities present in NFsim so that they can be integrated with other simulation frameworks, including MCell, we designed and implemented an API around the NFsim engine called *libNFsim*. *libNFsim* exposes the model specification setup and simulation functionality present in the NFsim suite as a set of library calls that can be integrated into third party simulation platforms as a shared library. We show in ***Fig. 2*** the methods available in the first release of the library, which is available as a standalone package at https://github.com/mcellteam/nfsimCInterface. The methods can be summarized as follows:

- **Model setup and initialization (*Fig. 3A*)**: The model is defined in an XML encoding of a BNGL file that is produced by BioNetGen and read by NFsim [16]. This specification is used to create the data structures that NFsim uses during the simulation, which include Parameters, Molecule Types, Reaction Rules and Observables. Once these data structures are created, the model is check-pointed such that the user can always reset to this point. After model setup, the model initialization defines the species copy numbers and reaction rate parameters.
- **Model simulation (*Fig. 3B,C*)**: *libNFsim* allows queries of reaction rule rates and control over specific rule firings. This fine-grained access to simulation functions is required for the integration with MCell, as discussed further below. For other applications, *libNFsim* allows calls to NFsim's standard simulation engine to propagate for a fixed number of steps or a fixed amount of simulation time.

- **Model state querying (*Fig. 3D*)**: *libNFsim* enables the user to query both static and dynamic properties of the model, such as the compartment structure, the copy number of species with specified properties, or the value of an arbitrary function of these properties.

*2.1.2 Handling of spatial compartments*

BioNetGen allows optional specification of compartmental information in a rule-based model [24]. Compartments are idealized, well-mixed spatial subvolumes that restrict how species can interact based on their location. A compartment is defined by its name, size, dimensionality and location in a compartment hierarchy defined by a tree structure. For example, ***Fig. 4*** shows a compartmental hierarchy composed of an extracellular container (EC), which surrounds a plasma membrane (PM) that encloses the cellular cytoplasm (CP). Note that in this hierarchy each three-dimensional (3D) container (EC or CP) can contain an arbitrary number of two-dimensional (2D) membrane structures (PM), whereas each 2D structure can enclose just a single 3D one. A species, which is comprised of molecules, can exist in either a single compartment (***Fig. 4A,C***) or can span multiple compartments if one of the constituent molecules is in a 2D compartment and is bound to molecules in one of the adjacent volume compartments (***Fig. 4B,D***).

Prior to the development of MCell-R, NFsim did not handle compartments in a BioNetGen model specification. Several situations that arise in spatial simulations, however, require tracking the location of constituent molecules in complexes. These include:

- Unbinding of volume-surface complexes: For example, in ***Fig. 4D***, the breaking of the bond between the molecules TF and R in the membrane-associated complexes should allow TF to return to the CP compartment, which is facilitated by associating the specific TF molecule with the compartment location, CP, as indicated by the '@CP' tag at the end of the BNGL string for the complex.
- Compartment aware reaction rules: The modeler may wish to restrict the spatial scenarios in which one or more reactants interact to produce products.

For this reason, a compartment attribute was added to the Molecule data structure in *libNFsim*, which communicates compartmental locations to MCell through the reaction graph strings used to identify a complex.

*2.1.3 Calculation of diffusion constants for complexes*

In a manually-specified reaction network, every species must be assigned a diffusion constant. In a rule-based model, however, because application of the rules may generate novel complexes that are not in the model specification, a method is required to assign diffusion constants to complexes based on their composition. The current implementation of MCell-R uses simple combining rules to determine the diffusion constants of complexes. For each uncomplexed molecule in the model specification (called a Molecule Type in BioNetGen), a default radius is specified. For molecules in a 3D volume, the molecular volume is assumed to be that of a sphere of the assigned radius. The radius of a complex is then taken as that of a sphere with a volume equal to the sum of the volumes of the constituent molecules. Similarly, for a surface molecule, the radius is used to calculate a corresponding surface area and the radius of a complex is determined by finding the radius of a circle with an area equal to the sum of areas of the constituent molecules. Mathematically, these combining rules can be written simply as

$$r_{complex} = \sqrt[3]{\sum_n r_n^3}$$
$$r_{surface} = \sqrt[2]{\sum_n r_n^2},$$

where $r_{complex}$ is the radius of a complex in a 3D volume, $r_{surface}$ is the radius of a surface complex, and the $r_n$ are the radii of the constituent molecules (3D or 2D as appropriate).

These complex radii are then converted to diffusion constants using either the Stokes-Einstein equation for 3D [37],

$$D = \frac{k_B T}{6\pi\eta r},$$

where $k_B$ is the Boltzmann Constant, *T* is temperature in Kelvin, $\eta$ is the viscosity and *r* is the sphere's Stokes radius, or the Saffman-Delbrück equation for 2D [38],

$$D = \frac{k_B T}{4\pi\mu h}\left(\log\frac{\mu h}{\eta r} - \gamma\right),$$

where $\mu$ is the viscosity of the membrane, *h* is the thickness of the membrane $\eta$ is the viscosity of the surrounding fluid, and $\gamma$ is the Euler constant.

*2.1.4 Extensions to the NFsim data model*

To enable the features defined in the preceding two sections as well as other integrations with MCell that are discussed below, we added to NFsim the concept of hierarchical namespaces: A namespace in this context is a set of properties associated to a single element in the BioNetGen object hierarchy. Consider for example the Stokes-Einstein formula for the diffusion of 3D complexes. The formula is a function of the Boltzmann constant (a system-wide property), the temperature (which can either be system-wide or compartment specific), the compartment's viscosity and the complex's Stokes radius, which is a function of the radii of its subunits. If we consider the diffusion of a particular complex it must have access to all the aforementioned variables. This can be solved if we consider a complex as a part of a hierarchy where it is contained by a compartment and the system-wide variables while at the same time being a container for its constituent subunits. Moreover, a complex has properties associated to itself like its diffusion function.

Implementation of hierarchical namespaces allows a given entity to access the variables associated with its containers and its subunits as required. Container relationships are dynamic and dependent on the state of the system. A property can be assigned to a given entity in the BNG-XML model specification as a *ListOfProperties* child entry associated with a Model, Compartment, Molecule Type, Reaction Rule, Species or another Property.

### 2.3 Extensions to MCell

#### *2.3.1 Modification of MCell's simulation algorithm to incorporate structured molecules*

To enable simulation of interactions between structured molecules and complexes, the core simulation algorithm of MCell was modified so that it queries *libNFsim* when specific information is required about possible reactions involving these species. These query points, indicated by the unshaded boxes in the schematic description of the MCell-R algorithm shown in ***Fig. 5***, occur following particle creation, particle collision, and reaction firing. In the remainder of this section we describe how events involving structured molecules are handled in greater detail.

**Particle creation**. In MCell when a new particle is created (“structured molecule is created” in ***Fig. 5***) either as a result of a reaction firing that creates new products or from a user-defined species release, MCell determines the set of unimolecular reactions it can undergo by comparing against a hash lookup table [14]. In MCell-R this set is determined by a call to *libNFsim* (“query NFsim for molecule properties”) that passes the graph pattern associated with the particle. *libNFsim* returns the propensities of the unimolecular reactions corresponding to the graph pattern, and the lifetime of the particle and the unimolecular process it undergoes at that time are chosen assuming that the firing times are exponentially distributed [14] (“lifetime calculation”). Unimolecular reactions are placed in a scheduling queue and fired at the appropriate time (“Unimolecular rule”).

**Particle collision**. *libNFsim* is invoked in a similar way following bimolecular collision events (“detect collision”) in MCell-R, which passes the graph patterns of both of the involved species (“query NFsim for bimolecular rules). *libNFsim* returns the propensities of the possible bimolecular reactions that can occur. MCell-R then determines whether a reaction occurs during the current time step and, if so, which of the possible biomolecular events occurs (“evaluate biomolecular reaction propensities”), according to previously described procedures [14], [36]. We note that using *libNFsim* to calculate the propensities of the colliding particles does not affect the accuracy of the simulation algorithm, but, in the case of species exhibiting a high degree of combinatorial complexity, may improve its efficiency.

**Reaction firing.** When MCell-R fires a reaction, either unimolecular or biomolecular, it queries *libNFsim* once again to obtain the graph structure of the corresponding products and associated properties, most notably the diffusion constant ("query NFsim for product molecules").

### *2.3.2 Extensions to the MCell data model and MDL*

Our implementation of MCell-R allows models to be constructed that involve both structured and unstructured molecules. The MCell data model has been extended in the following ways to distinguish between these types and also to include the necessary graph information about structured molecules that enables *libNFsim* to perform the necessary operations described in the preceding section. The primary extensions are

**Proxy molecule types**. In MCell-R a given molecule type can be marked with the *EXTERN* qualifier in the model definition, which instructs MCell to delegate all calculations about its reactions and properties to an external simulation engine (*libNFsim* in our implementation). All multi-component particles are then instances of these two base proxy types (see ***Note 1***).

**Graph patterns**. Proxy molecules are distinguished from each other by an associated graph pattern, stored as a string, which is used during communication with *libNFsim*. The graph string, which is created by *libNFsim* at initialization based on user input or when new particles are created by reaction rule firings, has a format similar to BNGL. The label for each distinct species is unique because the order of molecules and components is determined by the NAUTY graph labeling algorithm [39]. See ***Note 2*** for an example.

## 3. Methods

The current interface to MCell-R uses a hybrid language that extends the Model Description Language (MDL) of MCell with elements of the BioNetGen language (BNGL) to enable specification of multicomponent molecules and rules. The graphical interface for MCell, called CellBlender, will allow interactive specification of MCell-R models without requiring knowledge of MDLr (see ***Note 3***). In this section we describe the main elements of MDLr and present two models that demonstrate the basic capabilities of the MCell-R simulator. We conclude with a discussion and the current limitations and future plans.

### 3.1 Specifying an MCell-R model using MDLr

MDLr is an extension of the MDL language defined as a set of preprocessor directives that allow the user to introduce multi-state multi-component elements into an MCell model definition. The basic syntactic features are described below, but for more details about MDL see the MCell Quick Reference Guide and MCell Reaction Syntax documents available at http://mcell.org/documentation. The preprocessor is invoked on MDL sections that are preceded with the hash symbol. Sections have been modified to enable introduction of structured molecules, definition of reaction rules that operate on features of structured molecules, definition of a compartment hierarchy, release sites for structured molecules, and definition of output observables that track features of structured molecules and their complexes. Examples of each of these extensions are provided in the following subsections. A full grammar definition of the MDLr language extension is given in [40].

*3.1.1 Definition of molecule types*

Molecules are the basic building blocks of both MCell and BioNetGen models, but they have a different meaning in each that must be reconciled. In MCell, molecules represent the chemical species that function independently for the purpose of diffusion and or reaction. Thus, in MCell, a complex between two molecules, a ligand and a receptor for example, is represented as a distinct molecule. When a reaction occurs in MCell, the reactant molecules are deleted and replaced by the product molecule or molecules. In BioNetGen, molecules represent the building blocks of complexes. They may contain components that serve as binding sites to other molecules or that take on different states, which can represent covalent modification (e.g., phosphorylation) or conformations. When a reaction occurs in BioNetGen, the reacting molecules are transformed to match the product specification. For example, a ligand-receptor binding reaction may be carried out by adding a bond between components of a ligand molecule and a receptor molecule. In this way, BioNetGen tracks explicitly the binding and internal states of all species in the system, which include both individual molecules and complexes of molecules. MDLr expands the syntax for MCell molecule definition to include the BNGL syntax for defining structured molecules [20], [21], [41]. In this syntax, components are defined within parentheses and the allowed states of a component are defined by strings beginning with the '~' character. A pair of structured molecules representing ligand and receptor could be specified as follows in MDLr:

```
#DEFINE_MOLECULES {
    Lig(l,l){
        DIFFUSION_CONSTANT_3D = "Einstein_Stokes"
    }
    Rec(a,b~Y~pY,g~Y~pY){
```

```
            DIFFUSION_CONSTANT_2D = "Saffman_Delbruck"
        }
    }
```

The ligand molecule, 'Lig', has two identical components called 'l', and the receptor molecule, 'Rec', has three components, 'a', 'b', and 'g'. Both the 'b' and 'g' components have an associated state representing the unphosphorylated ('Y') and phosphorylated ('pY') state of tyrosine residues associated with specific receptor subunits. The diffusion constants associated with each of these molecule types are specified using the MDL keywords DIFFUSION_CONSTANT_3D and DIFFUSION_CONSTANT_2D, which identify the corresponding molecule types as volume and surface molecules respectively.

### *3.1.2 Definition of reactions*

Reactions in BioNetGen are generated by reaction rules the describe the properties that structure molecules must have in order to undergo a reaction and how the reaction transforms these molecules when it fires. To enable specification of rules, MDL's reaction syntax has been extended to allow BNGL-style rules to be entered in the DEFINE_REACTIONS block. The BNGL syntax was also slightly modified to require enclosing of rate constants in square brackets (the MDL convention). In addition, the rate constants must be specified in units following the MCell convention: $s^{-1}$ for unimolecular reactions, $M^{-1}\,s^{-1}$ for bimolecular volume reactions, and $\mu m^2\,s^{-1}$ for bimolecular surface reactions. An example of reaction specification in MDLr is as follows:

```
#DEFINE_REACTIONS{
    /* Ligand-receptor binding */
    Rec(a) + Lig(l,l) <-> Rec(a!1).Lig(l!1,l) [kp1, km1]
    /* Receptor-aggregation */
    Rec(a) + Lig(l,l!+) <-> Rec(a!2).Lig(l!2,l!+) [kp2, km2]
```

```
    /* Constitutive Lyn-receptor binding */
    Rec(b~Y) + Lyn(U,SH2) <-> Rec(b~Y!1).Lyn(U!1,SH2) [kpL, kmL]
}
```

The second rule provides an example of using a bond wildcard, ‘!+’, to specify binding of a receptor to a ligand molecule that is already bound at one of its l components. The third rule specifies the binding of a Lyn molecule to a Rec molecule at its b component, which must be both unbound and in the unphosphorylated (‘Y’) state.

### *3.1.3 Compartment hierarchy and molecule release*

The INSTANTIATE Scene command in MDL is used to define the compartments and the initial placement of molecules in the simulation. Compartments are defined using surface meshes, which are called OBJECTs in MDL. In MDLr, surface meshes must be closed and placed in a hierarchical structure corresponding to the specification of compartments in the INSTANTIATE Scene block. The MDL commands used to construct the actual triangulated surface meshes are omitted here for space reasons, but extensive examples and documentation can be found at the http://mcell.org web site, which also provides tutorials on mesh construction using the CellBlender GUI (see also [36]). The following MDLr code provides an example of compartment and molecule release site definition:

```
#INSTANTIATE Scene OBJECT {
  EC OBJECT EC {
    VISCOSITY = mu_EC
  }
  CP OBJECT CP {
    PARENT = EC
    VISCOSITY = mu_CP
    MEMBRANE = PM OBJECT CP[ALL]
    MEMBRANE_VISCOSITY = mu_PM
```

```
    }
    ligand_rel RELEASE_SITE{
      SHAPE = Scene.EC[ALL] - Scene.CP[ALL]
      MOLECULE = @EC:Lig(l,l)
      NUMBER_TO_RELEASE = Lig_tot
    }
    receptor_rel RELEASE_SITE{
      SHAPE = Scene.CP[PM]
      MOLECULE = @PM:Rec(a,b~Y,g~Y)
      NUMBER_TO_RELEASE = Rec_tot
    }
  }
```

Two nested volume compartments are defined here, the extracelluar compartment (EC) contains the cytoplasmic compartment (CP) with the plasma membrane surface compartment (PM) forming their boundary. The MCell OBJECTS EC and CP are both meshes whose geometry is defined in a separate file. In order to match BioNetGen's nested compartment hierarchy, MCellr extends mesh objects with the several attributes. The PARENT attribute defines the volume compartment inside which the current compartment resides. In this example, the parent compartment of CP is thus EC, whereas EC, because it is outermost in the hierarchy, does not have a PARENT. The MEMBRANE attribute defines the name of the surface compartment that forms the boundary between the current compartment and its parent. For CP, the MEMBRANE compartment is given the name PM. Naming membrane compartments is required for molecule placement in them. The string 'CP[ALL]' after OBJECT in the MEMBRANE definition defines the mesh elements that make up the surface compartment. In the current implementation of MCell-R surface compartments must be made up of closed meshes, so the membrane of any volume compartment must always be made up of the entire mesh that defines it.

In addition to these attributes, MDLr also allows the definition of viscosities associated with both the compartment and its associated membrane using the attributes VISCOSITY and MEMBRANE_VISCOSITY, as shown (note that the parameters following these declarations are defined elsewhere).

Particle placement is performed in MCell using the RELEASE_SITE object. Release sites are defined in the example above for both ligand and receptor molecules, 'ligand_rel' and 'receptor_rel' respectively. The SHAPE attribute is used to define the region into which particles will be released. Here, ligand molecules are released into the volume region between the mesh that defines EC ('Scene.EC[ALL]') and the mesh that defines CP ('Scene.CP[ALL]'). Receptor molecules, on the other hand, are released onto the surface mesh PM ('Scene.CP[PM]') (see ***Note 4***). In MDLr the allowed syntax of the MOLECULE attribute is extended to include BNGL specification of complexes. The compartmental attribute in the BNGL string, '@EC' for Lig molecules and '@PM' for Rec molecules, specifies the compartment over which molecules will be randomly placed. The number of molecules to be placed is set by the attributes NUMBER_TO_RELEASE.

*3.1.4 Specifying outputs*

In MDL the REACTION_DATA_OUTPUT command is used to define properties to track during a simulation. An example of such a property is the number of instances of species having a specified property, e.g., phosphorylation of a particular component or a bond between components of different molecules. In BioNetGen these outputs are called "Observables" and are specified using BNGL strings that may contain wildcards called "Patterns" [21], [41]. In MDLr, BNGL patterns may be used in the REACTION_DATA_OUTPUT block to specify outputs that are written to files during the simulation. Several examples are provided in the following MDLr code:

```
#REACTION_DATA_OUTPUT{
  STEP = 1e-6
  {COUNT[Rec(a!1).Lig(l!1,l), WORLD]} => "./react_data/RecMon.dat"
  {COUNT[Rec(a!1).Lig(l!1,l!2).Rec(a!2), WORLD]} => "./react_data/RecDim.dat"
  {COUNT[Lyn(U!1).Rec(b~Y!1,a), WORLD]} => "./react_data/LynRec.dat"
  {COUNT[Rec(b~pY!?), WORLD]} => "./react_data/RecPbeta.dat"
  …
}
```

The STEP keyword indicates the frequency (in seconds) at which observables are to be calculated and output to file. Each COUNT statement produces a count of the number of species in the simulation matching the specified pattern at each output time. The four patterns shown above correspond to the number of receptors bound to singly-bound ligands, the number of receptors bound to doubly-bound ligands (and hence in dimers), the number of Lyn molecules bound to unphosphorylated Rec molecules, and the number of Rec molecules that are phosphorylated on their b components respectively. For further details about BioNetGen pattern syntax see Ref. [21] and http://bionetgen.org. The WORLD keyword here as the second argument to COUNT indicates that species at any location are to be included in the count. It may be replaced by any valid MDL specification of a spatial region or mesh region, such as those discussed above in the definition of RELEASE_SITES. The arrow followed by a string indicates that the data at each output time is to be written to a file with the given path.

### 3.2 Examples and validation

We now present two examples of models that we have used to validate the correctness of our MCell-R implementation and also illustrate the types of biochemical complexity that can be naturally represented using rules. Full MDLr code for each of these examples is available at http://mcell.org.

### *3.2.1 Bivalent ligand bivalent receptor*

The bivalent ligand, bivalent receptor (BLBR) model [42] is a simple model of polymerization of cell surface receptors by a soluble ligand. This model tests the ability of MCell-R to handle simulation in a case where the network size is potentially very large. Indeed, the BLBR system can create polymer chains as long as the number of receptors in the system. A simple version of BLBR can be encoded by the following three BioNetGen rules:

```
L(r,r) + R(l) -> L(r!1,r).R(l!1) kp1 #Binding of free ligand
L(r,r!+) + R(l) -> L(r!1,r!+).R(l!2) kp2 #Cross-linking of ligand bound to receptor
L(r!1).R(l!1) -> L(r) + R(l) koff #Unbinding of ligand
```

The first rule describes the binding of free ligand from solution to a receptor. The requirement for free ligand is specified by the pattern 'L(r,r)', which requires two unbound r sites on the reacting ligand molecule. The second rule describes the binding of the second site on the bivalent ligand once the first site is bound. Here, the pattern 'L(r,r!+)' specifies an L molecule with one free site and one bound site (indicated by the wildcard '!+') as one of the reactants, and the unbound site is bound to the free receptor site specified by the reactant pattern 'R(l)'. The third rule specifies that dissociation of the ligand-receptor bond happens at the same rate regardless of whether the other site on the L molecule is bound. The full set of model parameters is shown in Table 1. For testing purposes, both ligand and receptor molecules are simulated as diffusing in a single volume compartment (CP).

To validate the accuracy of the simulations performed by MCell-R, we compared with results generated by execution of an equivalent reaction model using NFsim under conditions for which the well-mixed assumption is valid. Two thousand trajectories were generated using each simulator, and probability distributions were generated for two different observables, the number of doubly-bound ligands and the number of ligand-receptor bonds, at a range of simulation times (***Fig. 6***). We then applied the 2-sample Kolmogorov-Smirnoff (K-S) test and found that the resulting p-values had a mean of greater than 0.6 with a minimum value greater than 0.1, demonstrating that the results produced by the two simulators are statistically indistinguishable.

*3.2.2 The FcεRI signaling network*

The high-affinity receptor for Immunoglobulin E (IgE), known as FcεRI, plays a central role in inducing the inflammatory response of the immune system to allergens [43]. ***Figure 7*** presents the elements of an early rule-based model that was developed to describe the molecules and reaction events downstream of ligand engagement with this receptor [44], [45]. In this model, the receptor binds monovalently through its ‘a’ component, which represents the alpha subunit of the receptor complex, to a bivalent ligand, which represents a covalently cross-linked dimer of IgE molecules. The interactions in this model imply a large biochemical network containing 354 unique species and 3680 different reactions. This network is small enough to be generated in full by BioNetGen and simulated using MCell, which enables us to benchmark against simulations performed by MCell-R. This model is a good test of the spatial accuracy in the simulator given that it contains volume-surface and surface-surface reactions of varying time scales and in sufficient numbers. We simulated 2700 trajectories using both MCell and MCell-R versions of the model and computed probability distributions for different observables and time points as shown in ***Fig. 8***. As with the BLBR model, applying the K-S test to these distributions showed that the results of the two simulators are statistically equivalent.

### 3.3 Conclusions and outlook

In this chapter we have presented a spatial modeling framework that combines the particle-based reaction-diffusion simulation capabilities of MCell with a network-free approach to multi-state and multi-component molecules and complexes that enables simulation of systems exhibiting large scale combinatorial complexity. We tested and validated our framework with two systems that present combinatorial complexity: the bivalent ligand bivalent receptor (BLBR) system and the network of early events in FcεRI signaling. These are prototypes for many other cell regulatory networks in biology that exhibit combinatorial complexity and in which spatial effects may play an important role, including nephrin-Nck-N-Wasp signaling [46], aggregation of transmembrane adaptors in immunoreceptor signaling [47], and signaling in the postsynaptic density of neurons [32], [48], [49].

The development of the MCell-R framework required extensions to both NFsim and MCell, including the development of *libNFsim* as a general application programming interface (API) for network-free modeling capabilities and MDLr to incorporate rule-based elements into MCell's language for model specification. Central to the development of an efficient simulator was the modification of MCell's reaction-diffusion algorithm to obtain diffusion and reaction parameters based on the molecular composition of species.

The development of *libNFsim* also opens the door for the integration of the network-free framework with other platforms. For example, the WESTPA package [50] implements the weighted ensemble algorithm for the accurate and efficient sampling of rate events in models of complex dynamical systems. Although we have been able to integrate WESTPA with network-based modeling capabilities in BioNetGen and MCell [51], [52], the lack of a clear programming interface to NFsim has prevented integration of network-free capabilities, which will now be possible.

The current implementation of MCell-R has several limitations that need to be addressed in future versions of the software. The first limitation is revealed by our preliminary attempts to perform simulations of the trivalent ligand, bivalent receptor (TLBR) model, which is a simple extension of the BLBR model we used to validate the simulator above. For certain parameters, this model is known to exhibit a phase transition in which all receptor and ligand molecules in the system can form a single complex [53]. NFsim has been shown to perform accurate and efficient simulations in this region of parameter space [16], but the MCell-R version of this model does not produce accurate results under conditions in which large scale aggregates form (10s to 100s of molecules) unless the time step is set to an impractically small value (results not shown). The reason for this loss of simulation accuracy is that reaction rates increase with the number of ligand molecules in an aggregate, eventually becoming too large for any given choice of minimum time step in MCell. Identifying a robust solution to this issue that will preserve accuracy while not drastically increasing simulation time is a topic for future research. For now, we recommend that before performing simulations in MCell-R of systems where such phase transitions are possible, one first uses NFsim to simulate the system under well-mixed conditions to determine whether large scale aggregates are formed. These simulations can also be used to benchmark subsequent simulations in MCell-R.

Another issue that is inherent in the simulation of molecular complexes is that the MCell algorithm currently treats all complexes as point particles. Thus, effects like volume exclusion and the effect of complex structure on reactivity are not considered, limiting the accuracy of the resulting dynamics. We are currently working to extend the internal representation of complexes to incorporate 3D structure, which will affect both the particle movement and bimolecular reaction components of the MCell simulation algorithm.

Finally, the initial implementation of MCell-R lacks the visualization capabilities that are provided for MCell models by the CellBlender interface [36]. Work is currently under way to enable model specification, simulation, and visualization of MCell-R models using CellBlender and we encourage readers to check the MCell website for the latest information about software availability. Current plans include explicit rendering of the 3D structure of molecular complexes based on either default assumptions or user specifications. While such an approach enables fine-grain representation of complexes and sets the stage for modification of the simulation algorithm to use 3D structure to affect reactivity, a more coarse-grained visualization approach will also be required to visualize configurations with a large number of species. We plan to use graph patterns to that will alter glyph properties used to represent species, such as size and color. For example, glyph size might be tied to the number of molecules in a complex, and the number of specific molecule types or modifications might be used to set color and intensity. We anticipate that such visualization capabilities will facilitate analysis of spatial effects for many cell regulatory processes that are mediated by complex molecular interactions.

## Notes

1. MDL definitions for structured volume and surface molecules used in MCell-R interface have the additional keyword EXTERN to indicate that an external library is to be called to invoke specified functions on these molecule types. The MDL generated by the MDLr preprocessor is

```
DEFINE_MOLECULES
  {
      volume_proxy //proxy molecule type.
      {
          DIFFUSION_CONSTANT_3D = KB*T/(6*PI*mu_EC*Rs)
          EXTERN //new element
      }
      surface_proxy //proxy surface type.
      {
          DIFFUSION_CONSTANT_2D = KB*T*LOG((mu_PM*h/(Rc*(mu_EC+mu_CP)/2))-
gamma)/(4*PI*mu_PM*h)
          EXTERN //new element
      }
  }
```

2. The GRAPH_PATTERN keyword is used to add a graph label to a structured molecule definition in MDL. In this example, structured ligand molecules, are included in a RELEASE_SITE definition. The GRAPH_PATTERN keyword is followed by the NAUTY-ordered canonical representation of the BNGL string which is commented out on the line where the RELEASE_SITE is defined. This MDL code is generated automatically by the MDLr preprocessor.

```
INSTANTIATE Scene OBJECT
{
```

```
...
  Release_Site_s1 RELEASE_SITE //bng:@EC::Lig(l,l,s~Y)
  {
    SHAPE = Scene.EC[ALL] - Scene.CP[ALL]
    MOLECULE = volume_proxy
    NUMBER_TO_RELEASE = 50
    RELEASE_PROBABILITY = 1
    GRAPH_PATTERN = "c:l~NO_STATE!3,c:l~NO_STATE!3,c:s~Y!3,m:Lig@EC!0!2!1," //new element
  }
...
}
```

3. As of this writing, source code is available for prototype versions of MCell-R at the MCell repository on GitHub (https://github.com/mcellteam/mcell). See http://mcell.org for the latest availability and documentation.

4. In principle the SHAPE attribute is redundant when the compartment location of species to be placed is specified. We anticipate that future versions of MCell-R will perform particle placement without the need to explicitly define the SHAPE attribute, whose use will be reserved for situations where more specific control over the release location is desired.

## Acknowledgement

This work was supported in part by the US National Institutes of Health grants P41GM103712 and R01GM115805.

## References

[1] L. A. Chylek, L. A. Harris, C.-S. Tung, J. R. Faeder, C. F. Lopez, and W. S. Hlavacek, "Rule-based modeling: a computational approach for studying biomolecular site dynamics in

cell signaling systems.," *Wiley Interdiscip. Rev. Syst. Biol. Med.*, vol. 6, no. 1, pp. 13–36, Sep. 2013.

[2] D. T. Gillespie, "Exact stochastic simulation of coupled chemical reactions," *J. Phys. Chem.*, vol. 81, no. 25, pp. 2340–2361, Dec. 1977.

[3] A. Funahashi, Y. Matsuoka, A. Jouraku, M. Morohashi, N. Kikuchi, and H. Kitano, "CellDesigner 3.5: A Versatile Modeling Tool for Biochemical Networks," *Proc. IEEE*, vol. 96, no. 8, pp. 1254–1265, Aug. 2008.

[4] S. Hoops *et al.*, "COPASI--a COmplex PAthway SImulator," *Bioinformatics*, vol. 22, no. 24, pp. 3067–3074, Dec. 2006.

[5] T. M. Bartol *et al.*, "Computational reconstitution of spine calcium transients from individual proteins," *Front. Synaptic Neurosci.*, vol. 7, no. OCT, Oct. 2015.

[6] R. A. Kerr, H. Levine, T. J. Sejnowski, and W.-J. Rappel, "Division accuracy in a stochastic model of Min oscillations in Escherichia coli.," *Proc. Natl. Acad. Sci. U. S. A.*, vol. 103, no. 2, pp. 347–52, Jan. 2006.

[7] K. Takahashi, S. N. V. Arjunan, and M. Tomita, "Space in systems biology of signaling pathways - towards intracellular molecular crowding in silico," *FEBS Lett.*, vol. 579, no. 8, pp. 1783–1788, Mar. 2005.

[8] K. Takahashi, S. Tanase-Nicola, and P. R. ten Wolde, "Spatio-temporal correlations can drastically change the response of a MAPK pathway.," *Proc. Natl. Acad. Sci. U. S. A.*, vol. 107, no. 6, pp. 2473–8, Feb. 2010.

[9] I. I. Moraru *et al.*, "Virtual Cell modelling and simulation software environment.," *IET Syst. Biol.*, vol. 2, no. 5, pp. 352–362, Sep. 2008.

[10] J. Hattne, D. Fange, and J. Elf, “Stochastic reaction-diffusion simulation with MesoRD.,” *Bioinformatics*, vol. 21, no. 12, pp. 2923–2924, 2005.

[11] D. T. Gillespie, A. Hellander, and L. R. Petzold, “Perspective: Stochastic algorithms for chemical kinetics,” *J. Chem. Phys.*, vol. 138, no. 17, pp. 170901–144908, 2013.

[12] B. Drawert *et al.*, “Stochastic Simulation Service: Bridging the Gap between the Computational Expert and the Biologist,” *PLOS Comput. Biol.*, vol. 12, no. 12, p. e1005220, Dec. 2016.

[13] S. S. Andrews, N. J. Addy, R. Brent, and A. P. Arkin, “Detailed Simulations of Cell Biology with Smoldyn 2.1,” *PLoS Comput. Biol.*, vol. 6, no. 3, p. e1000705, Mar. 2010.

[14] R. A. Kerr *et al.*, “Fast Monte Carlo Simulation Methods for Biological Reaction-Diffusion Systems in Solution and on Surfaces,” *SIAM J. Sci. Comput.*, vol. 30, no. 6, pp. 3126–3149, Jan. 2008.

[15] W. S. Hlavacek, J. R. Faeder, M. L. Blinov, A. S. Perelson, and B. Goldstein, “The Complexity of Complexes in Signal Transduction,” *Biotechnol. Bioeng.*, vol. 84, no. 7, pp. 783–94, Dec. 2003.

[16] M. W. Sneddon, J. R. Faeder, and T. Emonet, “Efficient modeling, simulation and coarse-graining of biological complexity with NFsim.,” *Nat. Methods*, vol. 8, no. 2, pp. 177–183, Dec. 2011.

[17] M. L. Blinov, J. R. Faeder, B. Goldstein, and W. S. Hlavacek, “BioNetGen: software for rule-based modeling of signal transduction based on the interactions of molecular domains.,” *Bioinformatics*, vol. 20, no. 17, pp. 3289–3291, 2004.

[18] V. Danos, J. Feret, W. Fontana, and J. Krivine, “Scalable Simulation of Cellular Signaling

Networks," *Lect. Notes Comput. Sci.*, vol. 4807, pp. 139–157, 2007.

[19] M. Meier-Schellersheim, X. Xu, B. Angermann, E. J. Kunkel, T. Jin, and R. N. Germain, "Key role of local regulation in chemosensing revealed by a new molecular interaction-based modeling method," *PLoS Comput. Biol.*, vol. 2, pp. 0710–0724, 2006.

[20] L. A. Chylek, L. A. Harris, J. R. Faeder, and W. S. Hlavacek, "Modeling for (physical) biologists: an introduction to the rule-based approach.," *Phys. Biol.*, vol. 12, no. 4, p. 045007, Jul. 2015.

[21] J. R. Faeder, M. L. Blinov, and W. S. Hlavacek, "Rule-based modeling of biochemical systems with BioNetGen.," *Methods Mol. Biol.*, vol. 500, pp. 113–67, Jan. 2009.

[22] P. Boutillier *et al.*, "The Kappa platform for rule-based modeling," *Bioinformatics*, vol. 34, no. 13, pp. i583–i592, Jul. 2018.

[23] B. R. Angermann *et al.*, "Computational modeling of cellular signaling processes embedded into dynamic spatial contexts," *Nature Methods*, vol. 9. pp. 283–289, 2012.

[24] L. A. Harris, J. S. Hogg, and J. R. Faeder, "Compartmental rule-based modeling of biochemical systems," in *Proceedings of the 2009 Winter Simulation Conference (WSC)*, 2009, pp. 908–919.

[25] M. Lis, M. N. Artyomov, S. Devadas, and A. K. Chakraborty, "Efficient stochastic simulation of reaction–diffusion processes via direct compilation," *Bioinformatics*, vol. 25, no. 17, pp. 2289–2291, 2009.

[26] O. Sorokina, A. Sorokin, J. D. Armstrong, and V. Danos, "A simulator for spatially extended kappa models," *Bioinformatics*, vol. 29, no. 23, pp. 3105–3106, 2013.

[27] S. S. Andrews, "Smoldyn: Particle-based simulation with rule-based modeling, improved

molecular interaction and a library interface," *Bioinformatics*, vol. 33, no. 5, pp. 710–717, Mar. 2017.

[28] P. J. Michalski and L. M. Loew, "SpringSaLaD: A Spatial, Particle-Based Biochemical Simulation Platform with Excluded Volume.," *Biophys. J.*, vol. 110, no. 3, pp. 523–529, Feb. 2016.

[29] G. Grünert, B. Ibrahim, T. Lenser, M. Lohel, T. Hinze, and P. Dittrich, "Rule-based spatial modeling with diffusing, geometrically constrained molecules," *BMC Bioinformatics*, vol. 11, no. 1, p. 307, 2010.

[30] G. Grünert and P. Dittrich, "Using the SRSim Software for Spatial and Rule-Based Modeling of Combinatorially Complex Biochemical Reaction Systems," vol. 6501, pp. 240–256, 2011.

[31] R. Suderman, E. D. Mitra, Y. T. Lin, K. E. Erickson, S. Feng, and W. S. Hlavacek, "Generalizing Gillespie's Direct Method to Enable Network-Free Simulations," *Bull. Math. Biol.*, pp. 1–27, Mar. 2018.

[32] P. J. Michalski and L. M. Loew, "CaMKII activation and dynamics are independent of the holoenzyme structure: an infinite subunit holoenzyme approximation.," *Phys. Biol.*, vol. 9, no. 3, p. 036010, 2012.

[33] J. S. Hogg, L. A. Harris, L. J. Stover, N. S. Nair, and J. R. Faeder, "Exact hybrid particle/population simulation of rule-based models of biochemical systems.," *PLoS Comput. Biol.*, vol. 10, no. 4, p. e1003544, Apr. 2014.

[34] N. Le Novère and T. S. Shimizu, "STOCHSIM: modelling of stochastic biomolecular processes.," *Bioinformatics (Oxford, England)*, vol. 17, no. 6. pp. 575–6, 2001.

[35] J. Colvin, M. I. Monine, R. N. Gutenkunst, W. S. Hlavacek, D. D. Von Hoff, and R. G. Posner, “RuleMonkey: software for stochastic simulation of rule-based models.,” *BMC Bioinformatics*, vol. 11, p. 404, Jan. 2010.

[36] S. Gupta *et al.*, “Spatial Stochastic Modeling with MCell and CellBlender,” in *Quantitative Biology: Theory, Computational Methods and Examples of Models*, B. Munksy, W. Hlavacek, and L. Tsimring, Eds. Cambridge, MA: MIT Press, 2018.

[37] C. C. Miller, “The Stokes-Einstein Law for Diffusion in Solution,” *Proc. R. Soc. London. Ser. A, Contain. Pap. a Math. Phys. Character*, vol. 106, no. 740, pp. 724–749, 1924.

[38] P. G. Saffman and M. Delbrück, “Brownian motion in biological membranes.,” *Proc. Natl. Acad. Sci. U. S. A.*, vol. 72, no. 8, pp. 3111–3, Aug. 1975.

[39] B. D. McKay, “Practical graph isomorphism,” *Congr. Numer.*, vol. 30, pp. 45–87, 1981.

[40] J. J. Tapia, “A study on systems modeling frameworks and their interoperability,” University of Pittsburgh, 2016.

[41] J. A. Sekar and J. R. Faeder, “Rule-based modeling of signal transduction: a primer.,” *Methods Mol. Biol.*, vol. 880, 2012.

[42] A. S. Perelson and C. DeLisi, “Receptor clustering on a cell surface. I. theory of receptor cross-linking by ligands bearing two chemically identical functional groups,” *Math. Biosci.*, vol. 48, no. 1–2, pp. 71–110, Feb. 1980.

[43] A. M. Gilfillan and J. Rivera, “The tyrosine kinase network regulating mast cell activation,” *Immunol. Rev.*, vol. 228, no. 1, pp. 149–169, Mar. 2009.

[44] B. Goldstein, J. R. Faeder, W. S. Hlavacek, M. L. Blinov, A. Redondo, and C. Wofsy, “Modeling the early signaling events mediated by FcepsilonRI.,” in *Molecular*

*Immunology*, 2002, vol. 38, no. 16–18, pp. 1213–1219.

[45] J. R. Faeder *et al.*, "Investigation of early events in Fc epsilon RI-mediated signaling using a detailed mathematical model.," *J. Immunol.*, vol. 170, pp. 3769–3781, 2003.

[46] C. V. Falkenberg, M. L. Blinov, E. U. Azeloglu, S. R. Neves, R. Iyengar, and L. M. Loew, "A Mathematical Model for Nephrin Localization in Podocyte Foot Processes," *Biophys. J.*, vol. 102, no. 3, p. 593a–594a, Jan. 2012.

[47] A. Nag, M. I. Monine, J. R. Faeder, and B. Goldstein, "Aggregation of membrane proteins by cytosolic cross-linkers: theory and simulation of the LAT-Grb2-SOS1 system.," *Biophys. J.*, vol. 96, no. 7, pp. 2604–23, Apr. 2009.

[48] M. I. Stefan, T. M. Bartol, T. J. Sejnowski, and M. B. Kennedy, "Multi-state Modeling of Biomolecules," *PLoS Comput. Biol.*, vol. 10, no. 9, 2014.

[49] P. J. Michalski, "The delicate bistability of CaMKII.," *Biophys. J.*, vol. 105, no. 3, pp. 794–806, Aug. 2013.

[50] M. C. Zwier *et al.*, "WESTPA: An Interoperable, Highly Scalable Software Package for Weighted Ensemble Simulation and Analysis," *J. Chem. Theory Comput.*, vol. 11, no. 2, pp. 800–809, Feb. 2015.

[51] R. M. Donovan, A. J. Sedgewick, J. R. Faeder, and D. M. Zuckerman, "Efficient stochastic simulation of chemical kinetics networks using a weighted ensemble of trajectories.," *J. Chem. Phys.*, vol. 139, no. 11, p. 115105, Sep. 2013.

[52] R. M. Donovan *et al.*, "Unbiased Rare Event Sampling in Spatial Stochastic Systems Biology Models Using a Weighted Ensemble of Trajectories," *PLOS Comput. Biol.*, vol. 12, no. 2, p. e1004611, Feb. 2016.

[53] B. Goldstein and A. S. Perelson, "Equilibrium theory for the clustering of bivalent cell surface receptors by trivalent ligands. Application to histamine release from basophils.," *Biophys. J.*, vol. 45, no. 6, pp. 1109–1123, 1984.

[54] J. R. Faeder *et al.*, "Investigation of early events in FcεRI-mediated signaling using a detailed mathematical model.," *J. Immunol.*, vol. 170, no. 7, pp. 3769–3781, 2003.

[55] W. Xu, A. M. Smith, J. R. Faeder, and G. E. Marai, "RULEBENDER: A Visual Interface for Rule-Based Modeling.," *Bioinformatics*, Apr. 2011.

[56] A. M. Smith, W. Xu, Y. Sun, J. R. Faeder, and G. E. Marai, "RuleBender: integrated modeling, simulation and visualization for rule-based intracellular biochemistry," *BMC Bioinformatics*, vol. 13, no. Suppl 8, p. S3, 2012.

[57] J. A. P. Sekar, J.-J. Tapia, and J. R. Faeder, "Automated visualization of rule-based models," *PLOS Comput. Biol.*, vol. 13, no. 11, p. e1005857, Nov. 2017.

## Tables

**Table 1**. Parameters for the BLBR model**.**

| Category | Parameter | Description | Value |
|---|---|---|---|
| *Initial populations* | `L0` | Number of ligand molecules | 5973 |
| | `R0` | Number of receptor molecules | 300 |
| *Reaction Rates* | `kp1` | Free ligand binding rate constant | 1.084e6 $M^{-1}s^{-1}$ |
| | `kp2` | Ligand cross-linking rate constant | 3.372e8 $M^{-1}s^{-1}$ |
| | `koff` | Ligand-receptor unbinding rate constant | 0.01 $s^{-1}$ |
| *Spatial parameters* | `Vol_CP` | CP volume | 39 µm |
| | `D_3D` | Diffusion constant for volume molecules | $10^{-4}$ $cm^2/s$ |

## Figure Captions

**Figure 1**. Overview of MCell-R model specification and simulation. The model is defined in an MDLr file and processed by the MDLr preprocessor, which generates input files for both MCell (MDL) and BioNetGen (BNGL) that are then used to initialize MCell and *libNFsim* respectively.

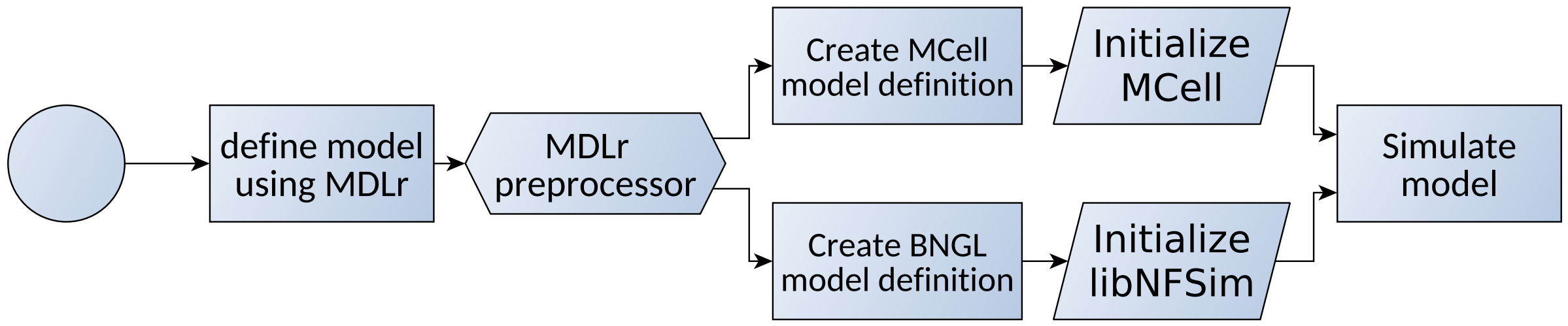

**Figure 2**. Functions available in the *libNFsim* API classified by their functionality. *Model setup* methods provide *libNFsim* with the basic model definition and checkpointing functionalities. *Model initialization* methods allow the user to set the initial species copy numbers and model parameter values. The *Experiment setup* methods allow the user to specify a full simulation protocol that may start and stop the simulation, change model parameters or species concentrations, and query different model observables.

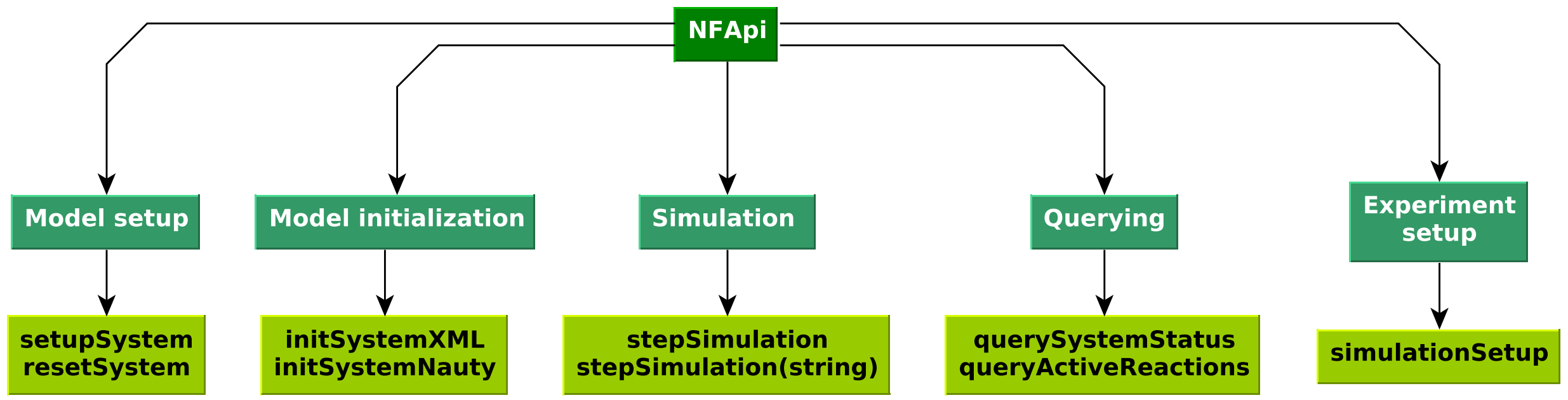

**Figure 3**. Simulation workflow with *libNFsim*. The API functions are represented as arrows with the inputs and outputs shown as tables. **A**) Simulation initialization is done by providing a list of reactant species, specified here using BNGL strings, and copy numbers. In this example, the first BNGL string specifies a ligand-receptor complex is initially present. It's overall compartment location is PM, as indicated by the initial '@PM', but the L molecule resides in the EC and upon dissociation from the receptor complex the L molecule would be a species located in EC. **B**) A call to the query function 'queryActiveReactions' returns a list of reactions rules with non-zero propensities. **C**) The 'stepSimulation' function fires a specific reaction rule. This function is used by MCell-R, which uses an MCell function to select over the active reactions when a species-species collision occurs. **D**) The 'queryStatus' function can be used to determine the number of species matching a particular query pattern. Here, the number of receptor-ligand complexes with two ligand molecules is returned.

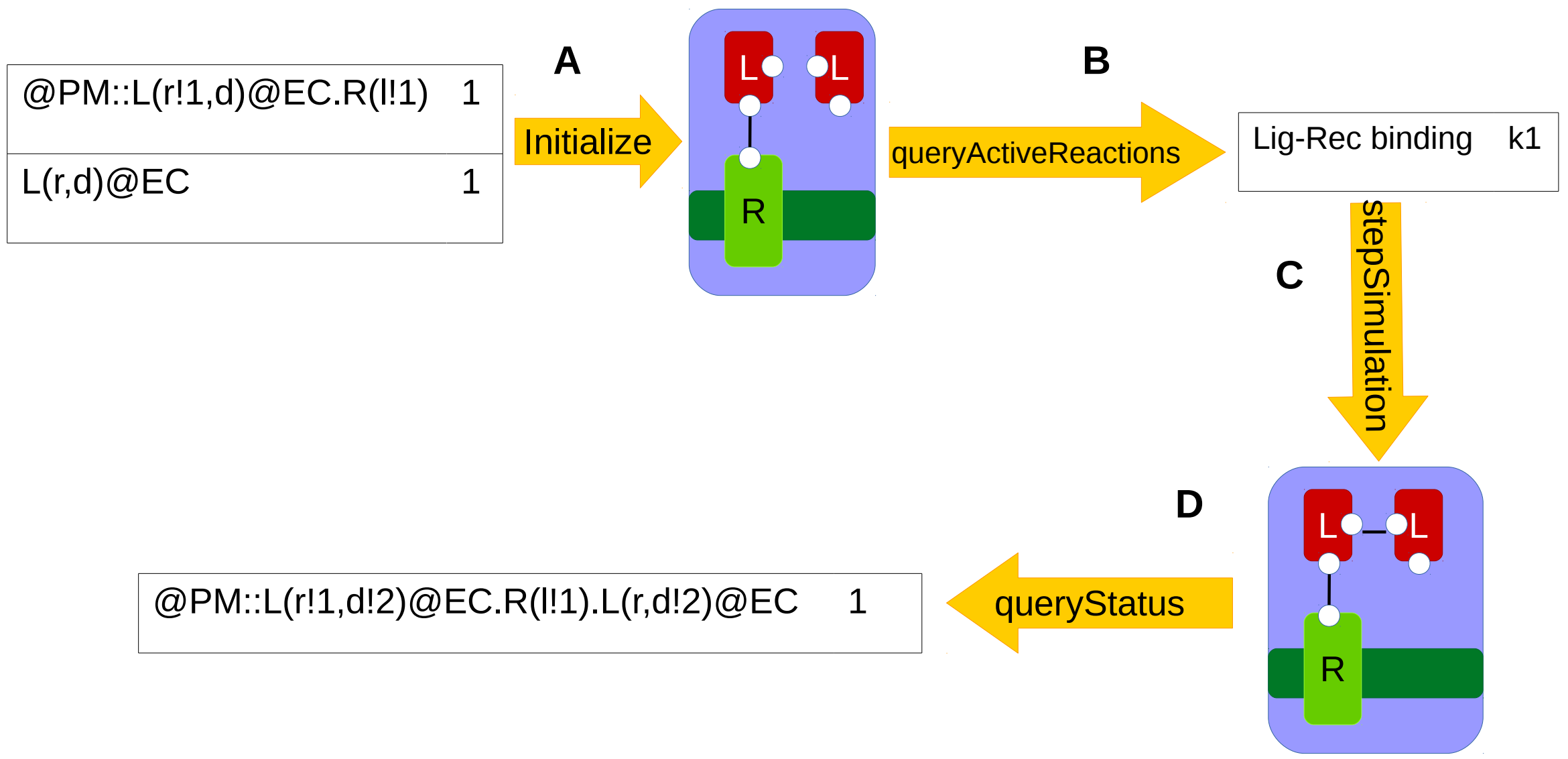

**Figure 4**. Representation of multi-state molecules and complexes using compartmental BioNetGen language (from [24] – copyright permission will be requested or the figure will be redrawn). (i) Ligand dimer located in EC. (ii) Tetrameric complex consisting of two L and two R molecules. The species is localized to the PM by the Rs, but the L molecules remain in the EC. (iii) TF dimers localized to CP. The dimerization component, 'd', must be in the state 'pY' for the bond to form. (iv) TF bound to a phosphorylated R molecule in the PM from the adjacent CP compartment.

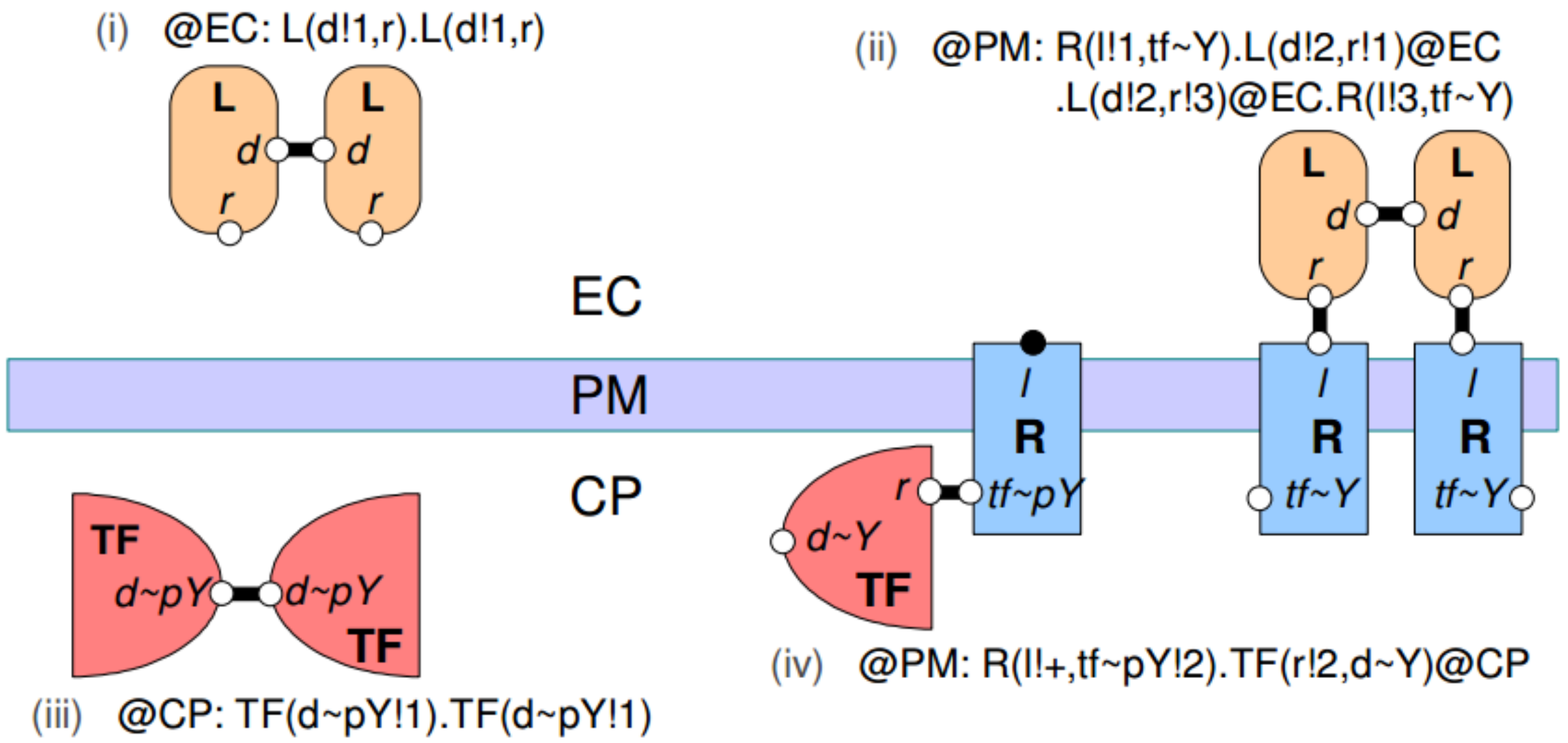

**Figure 5**. MCell simulation algorithm with modifications for MCell-R. The boxes with a white background indicate the points at which MCell calls functions in the *libNFsim* API.

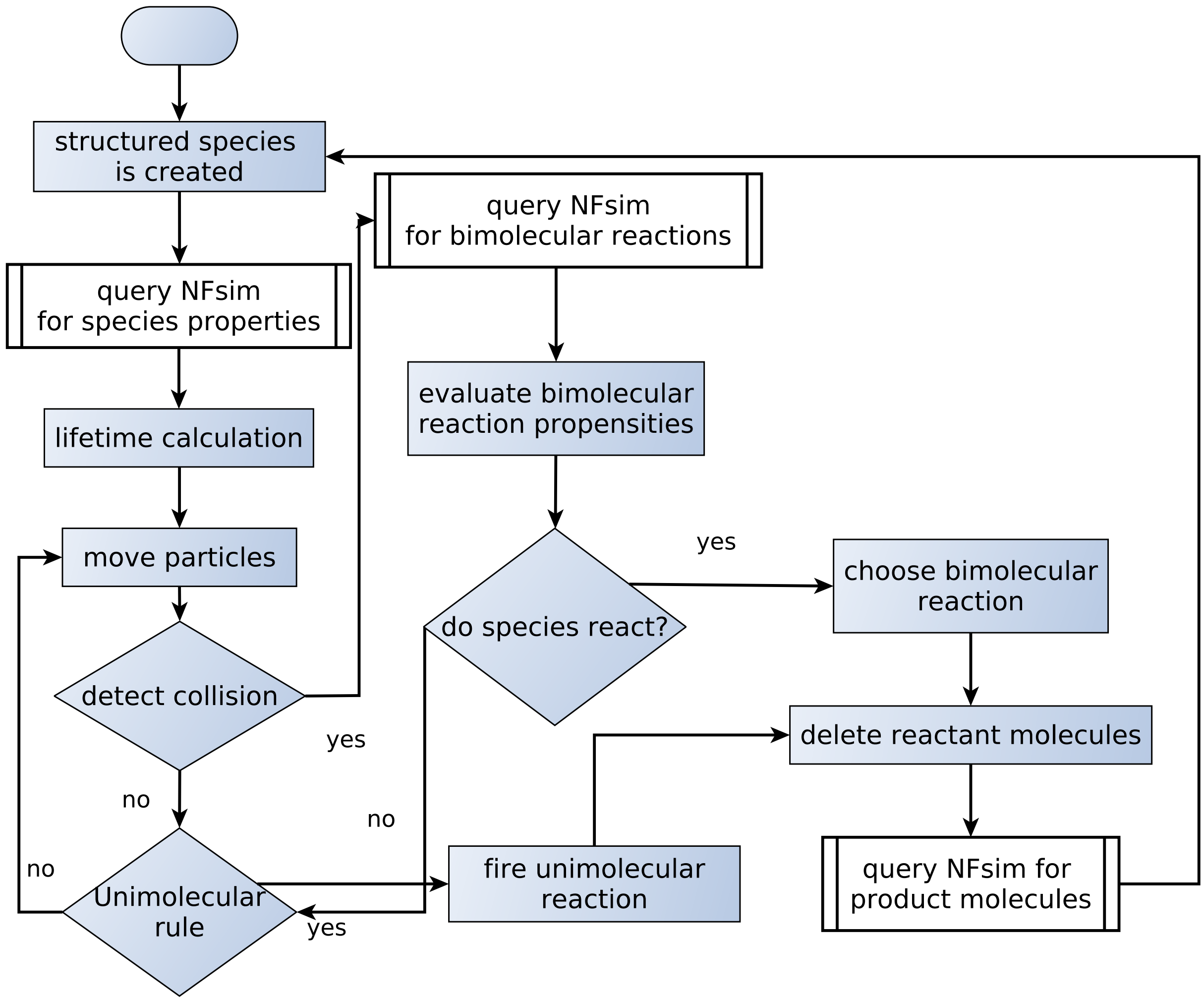

**Figure 6**. Validation of MCell-R simulation results for the BLBR model. Probability distributions are shown for two observables (columns) at five different time points (rows) computed from 2000 simulations using either MCell-R (green lines) or NFsim with the well-mixed version of the model (blue lines). Applying a two-sample Kolmogorov-Smirnoff test over the set of distributions confirms the accuracy of the results because no statistically significant differences are observed between the distributions.

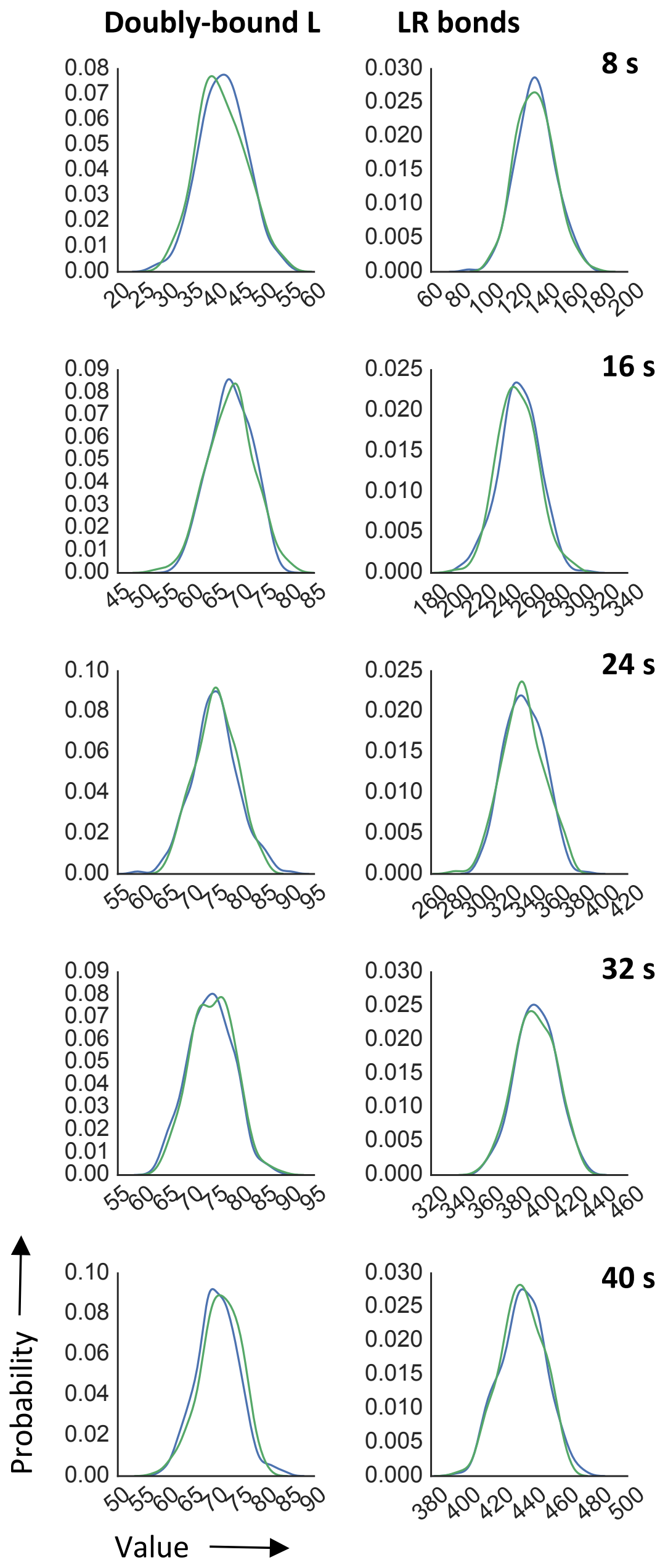

**Figure 7**. Graphical representations of the model of early events in FcεRI signaling from [44], [54]. **A**) Contact map of the model, which includes four molecule types—Lig, Rec, Lyn, And Syk—as rendered by RuleBender, a graphical interface for BioNetGen models [55], [56]. Components with purple background indicate the presence of multiple component states,

which in this model represent phosphorylation. **B**) Bipartite representation of the model using the atom-rule graph defined in Ref. [57]. Unbound components, component states, and bonds comprise one type of node (shaded pink) in the graph, and rules (shaded purple) comprise the other. Darker edges are used to indicate nodes that are consumed (outgoing) or produced (incoming) by the corresponding rule. Lighter edges indicate nodes that are required for the corresponding rule to fire. The rule nodes labeled 'RG' correspond to groups of nodes that have the same effect but may have different requirements.

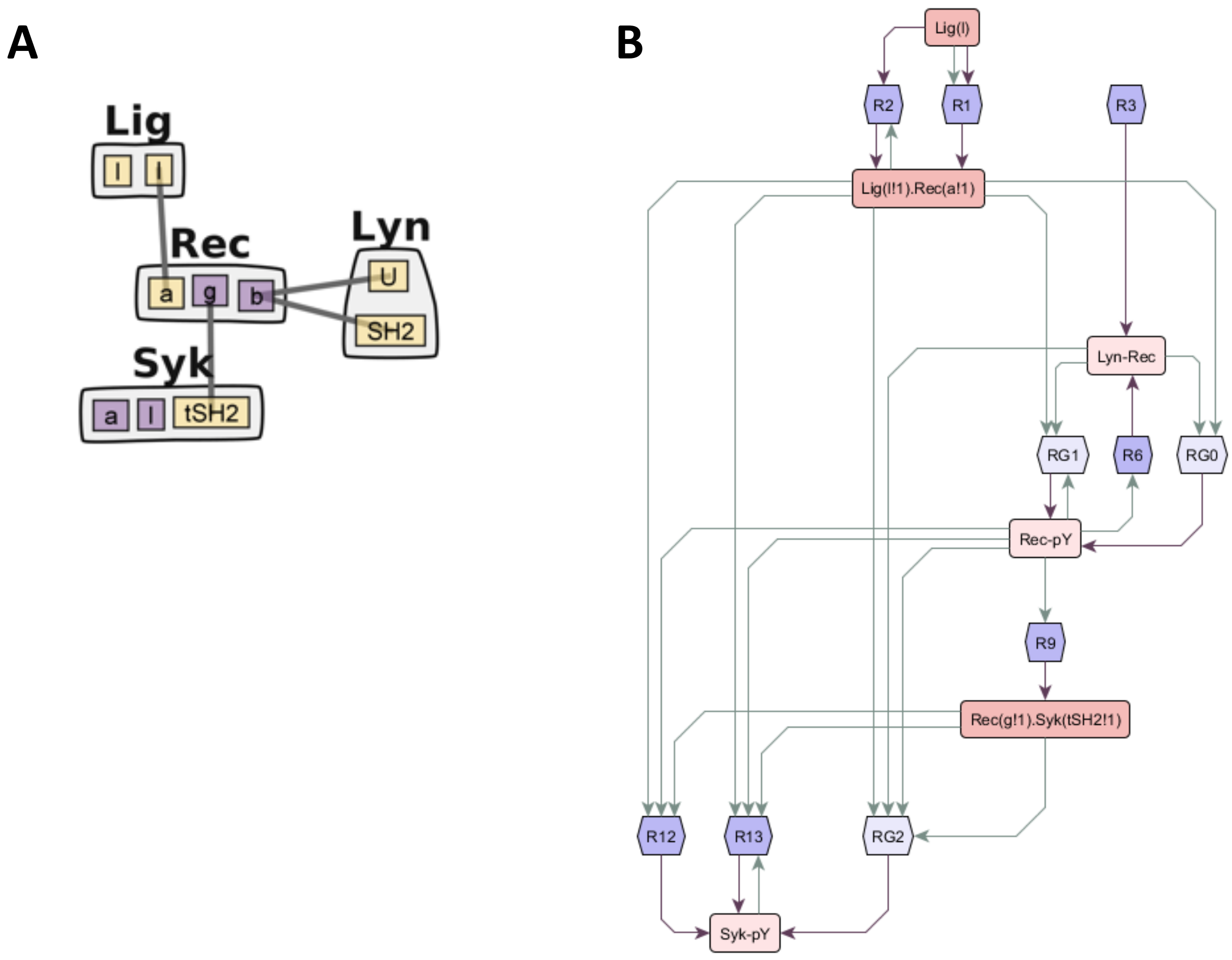

**Figure 8**. Validation of MCell-R simulation results for the FcεRI model. Probability distributions are shown for five observables (columns) at five different time points (rows)

computed from 2700 simulations using either MCell-R (green lines) or MCell with a pre-generated reaction network (blue lines). Applying a two-sample Kolmogorov-Smirnoff test over the set of distributions confirms the accuracy of the results because no statistically significant differences are observed between the distributions.

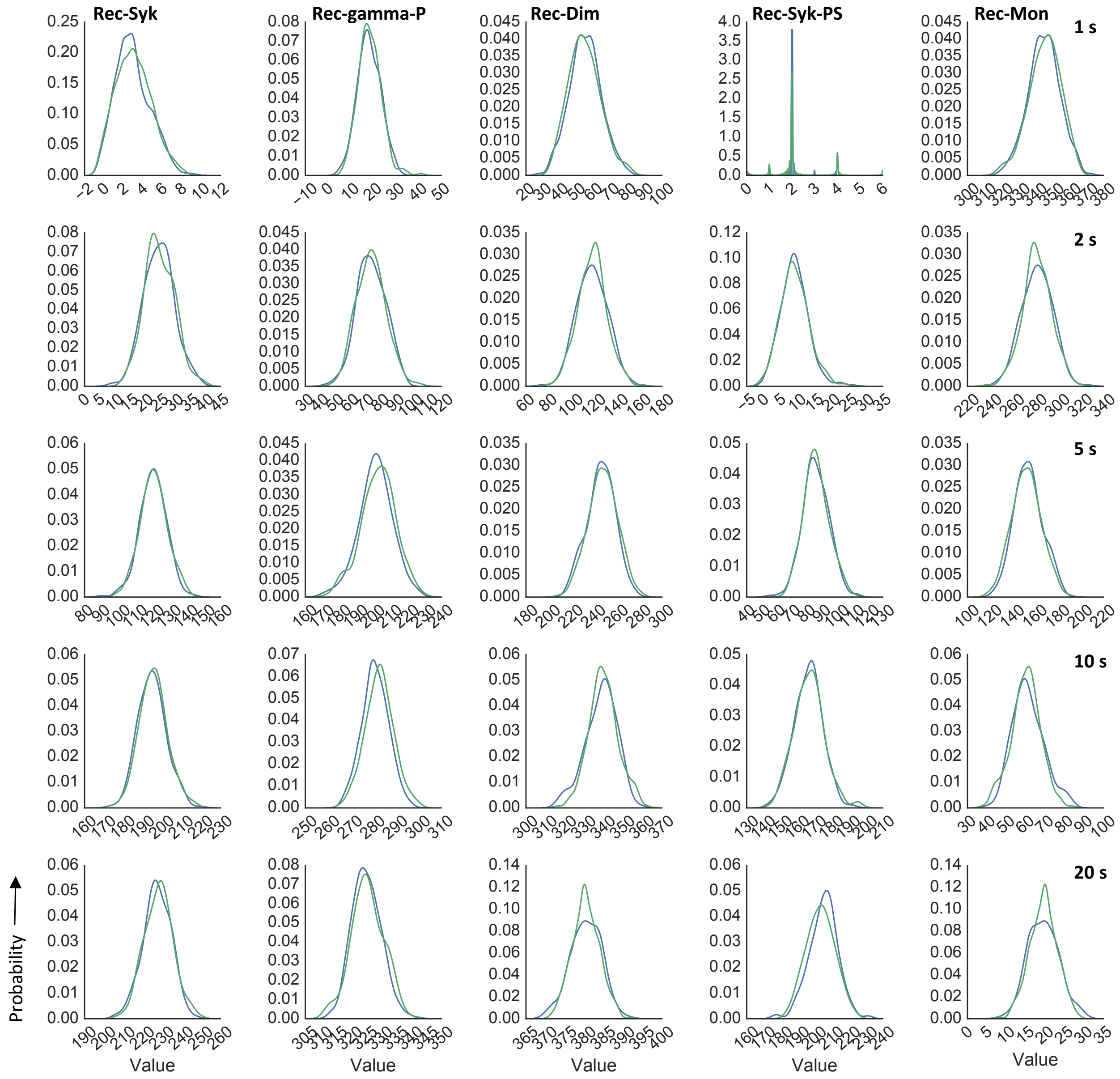